\documentstyle[preprint,aps]{revtex}
\input epsf

\begin{document}
\draft
\title{Higher-Order Neural Networks, Poly\`a Polynomials,
and Fermi Cluster Diagrams }

\author{K. E. K\"urten}
\address{Institut f\"ur Experimentalphysik, Universit\"at Wien, Austria}
\author{J. W. Clark}
\address{McDonnell Center for the Space Sciences and Department of Physics,\\
	 Washington University, St. Louis, Mo 63130, USA}

\maketitle
\maketitle
\begin{abstract}

The problem of controlling higher-order interactions in neural
networks is addressed with techniques commonly applied in
the cluster analysis of quantum many-particle systems.
For multi-neuron synaptic weights chosen according to
a straightforward extension of the standard Hebbian
learning rule, we show that higher-order contributions
to the stimulus felt by a given neuron can be readily evaluated
via Poly\`a's combinatoric group-theoretical approach or
equivalently by exploiting a precise formal analogy 
with fermion diagrammatics.

\end{abstract}

In attempting to unravel the mechanisms of information processing
and attendant adaptive behavior in neurobiological systems,
considerable attention is currently being directed to
non-linear processing in dendritic trees and to the computational
power that can be gained from multiplicative or higher-order
interactions between neurons \cite{dendrites,mel}.  This focus
is supported by a large body of theoretical work demonstrating
enhanced performance in artificial neural networks involving
such higher-order or multi-neuron interactions, as applied
to a variety of information-processing tasks, most notably
memory storage and recall 
\cite{lee,peretto,sompolinsky,baldi,personnaz,feigelman,gardner,abbott,horn,psaltis,arenzon}.  Introduction of higher-order
couplings is accompanied, however, by the threat of a combinatoric
explosion that may strongly inhibit analysis, evaluation, and
optimization.  In this note we expose some simple techniques based 
on group-theoretic symmetry arguments that serve, in some cases,
to reduce the serverity of these problems and give access to 
the advantages of higher-order networks for problem domains 
involving complex correlations.  Our study is guided by interesting
parallels with the diagrammatic analysis of fermion clusters in 
many-body physics. 

We consider the following simple but standard model of a higher-order 
neural network.  The network consists of $N$ binary-output hard-threshold 
units (model neurons) $i$ whose state variables $\sigma_i$ take the 
value $+1$ if the unit is active (``firing'') and $-1$ if the unit is 
inactive (``not firing'').  Model neuron $i$ receives inputs from 
exactly $K_i$ other units of the network, with self interactions
excluded so that $1 \leq K_i \leq N-1$.  A given neuron updates
its state on a discrete time grid according to the deterministic 
threshold rule
\begin{equation}
\sigma_i(t+1) = {\rm sgn} \left[ h_i(t) \right]\,, \qquad i=1,\ldots,N \,.
\label{dynamics}
\end{equation}
Here $h_i(t)$ is the net stimulus felt by the neuron at time $t$,
coming from internal and external inputs but reduced by a threshold
parameter.  For our purposes it is immaterial whether sequential
or parallel updating is imposed.  The general higher-order synaptic 
structure of the network model is expressed in the assumed form
\newpage
\begin{eqnarray}
h_i(t)&=& c_{i0}(t)+ \sum_{j_1}c_{ij_1}(t)\sigma_{j_1} + 
+ \sum_{j_1j_2}c_{ij_1j_2}(t)\sigma_{j_1}(t)\sigma_{j_2}(t) + \cdots +
\nonumber \\
&\qquad+& \sum_{j_1 < j_2 < \cdots < j_{K_i}} c_{ij_1j_2\cdots j_{K_i}}(t) 
\sigma_{j_1}(t) \sigma_{j_2}(t) \cdots \sigma_{j_{K_i}}(t) \nonumber \\
&=& C_0(t) + C_1(t) + C_2(t) + \cdots + C_{K_i}(t) \, ,
\label{netstimulus}
\end{eqnarray}
where the sums include only those $K_i$ neurons from which
neuron $i$ receives inputs.  The first term represents any
external input to neuron $i$ (reduced by its threshold), while
the second term is the usual one representing binary interactions,
a simple linear sum of states of input neurons weighted by synaptic strengths
$c_{ij_1}$.  The higher-order terms in the expansion, for $n \geq 2$,
represent ``multiplicative'' interactions in that they are
linear combinations of the {\it products} of two or more
input-neuron states.  One also speaks of a ``sum-of-products''
form for such interactions.

We observe that the general $n$th-order contribution, 
\begin{equation}
C_n = \sum_{j_1<j_2\cdots < j_n} c_{ij_1j_2 \cdots j_n} \sigma_{j_1}
\sigma_{j_2}\cdots \sigma_{j_n}\,,
\label{generic}
\end{equation}
representing the irreducible interaction of $n$ neurons with 
neuron $i$, introduces ${K_i \choose n} = K_i!/n!(K_i-n)!$ weight
parameters.  Accordingly, specification of the net stimulus
(\ref{netstimulus}) requires $2^{K_i}$ parameters.  The 
exponential explosion of parameters with increasing connectivity
$K_i$ has deterred widespread application of higher-order networks, 
in spite of their theoretical advantages.

Indeed, complete optimization of a network of a network having 
all possible combinations of higher-order terms is patently
impractical for sizable values of $K_i$ typically needed in 
real-world applications.  However, a restricted optimization 
problem has been attacked by retaining only a strongly reduced 
pattern-specific connectivity \cite{kuerten1,kuerten2}, while 
otherwise implementing the extended Hebbian learning rule to 
be introduced below.  A similar strategy based on a 
connection-pruning scheme adapted to the pattern domain has 
been employed to tame the combinatoric explosion of parameters 
in higher-order probabilistic perceptrons \cite{clark}.

Of course, if the entire array of coefficients $c_{ij_1j_2 \cdots j_n}$
is specified at the outset, the explosive combinatoric optimization
problem becomes moot.  In this note we shall focus on the fully connected
network in an important special case of ``one-shot'' learning
in which it is feasible and straightforward to evaluate the 
general term $C_n$ of the series (\ref{netstimulus}).  In fact,
by exploiting standard group-theoretic results, we are
actually able to sum this series in the limit of asymptotically 
large connectivity ($K_i \to \infty$, implying an infinitely 
large network). 

We consider the familiar task of storage and recall of $p$
random patterns $S^{\mu} = \{ S_1^\mu,S_2^\mu,\ldots,S_2^N\}$
in the firing activities of the neuronal units, where again
$S_j\in\{-1,1\}$.  As is well known \cite{peretto,personnaz,feigelman},
such patterns can be faithfully stored as fixed points of the
dynamics (\ref{dynamics}) of the network model to a capacity 
$p=O(N^K)$ (with $K = {\rm min}_i K_i$), if the weight parameters of 
the stimulus expression (\ref{netstimulus}) are chosen according 
to an extension of the classical Hebbian learning rule to the presence of
interactions of all orders up to $K_i$:
\begin{equation}
c_{ij_1j_2\cdots j_n}= \sum_{\mu=1}^p S_i^\mu S_{j_1}^\mu
S_{j_2}^\mu \cdots S_{j_n}^\mu \,, \quad n=1,\ldots,K_i \, .
\label{genHebb}
\end{equation}
The efficacy of memory storage is commonly analyzed in terms
of the overlaps 
\begin{equation}
m^\mu (t)= \sum_{j}S^\mu_j \sigma_j (t) 
\end{equation}
of the current network configuration $\{\sigma_1(t),\sigma_2(t),\cdots,
\sigma_N(t)\}$ with a given pattern $S^\mu$.  When a relative-entropy 
cost function is adopted \cite{qian}, this specification
can be shown to be optimal among the class of simple local learning
rules (where ``local'' implies that changes of synaptic strength 
depend only on the states of the neurons interacting at the 
given synapse).

The generic term (\ref{generic}) in the stimulus expansion
(\ref{netstimulus}) is evaluated as follows.  We first examine 
the modified $n$th-order contribution 
\begin{equation}
  {\overline C}_n =   \sum_{j_1...j_n} c_{ij_1...j_n}
 \sigma_{j_1}(t)...\sigma_{j_n}(t)  
\label{auxiliary}
\end{equation}
to the net stimulus, which consists of $K_i^n$ terms.
This auxiliary quantity contains redundant terms of
two kinds: (i) ``diagonal'' terms in which two or more
of the indices $j_1,\ldots,j_n$ coincide and (ii) ``symmetrical''
terms differing only through a permutation of distinct labels
$j_1,\ldots,j_n$, which may be combined into a single
term by redefining the weight parameter $c_{j_1\cdots j_n}$
as the sum of the weight parameters with permuted indices.
The former terms are redundant because they already appear in
lower-order contributions of the expansion (\ref{netstimulus}).
The latter terms lead to overcounting by a factor $n!$.

Inserting the learning rule (\ref{genHebb}) into Eq.~(\ref{auxiliary})
and interchanging the order of the summations, we may write 
\begin{eqnarray}
     {\overline C}_n &=&  \sum_{j_1...j_n} c_{ij_1\cdots j_n}
 \sigma_{j_1}\cdots \sigma_{j_n} \ = \sum_{j_1\cdots j_n}\sum_{\mu=1}^p
 S_i^\mu S_{j_1}^\mu \cdots S_{j_n}^\mu \sigma_{j_1}...\sigma_{j_n}
   \nonumber \\
&=& \quad \sum_{\mu=1}^p S_i^\mu \left( \sum_{j_1} 
S_{j_1}^\mu \sigma_{j_1}\right)\cdots\left( \sum_{j_n}
 S_{j_n}^\mu \sigma_{j_n}\right) =
 \sum_{\mu=1}^p S_i^\mu\left[\sum_j S_j^\mu \sigma_j\right]^n
 \, .  
\end{eqnarray}
The desired $n$th-order contribution $C_n$ and its modified counterpart
${\overline C}_n$ are evidently related by
\begin{equation}
 C_n = \sum_{j_1<...<j_n} c_{ij_1...j_n}
  \sigma_{j_1}... \sigma_{j_n} = {1 \over n!} \overline C_n
{\rm det} (\delta_{j_\alpha j_\beta}) \, .
\label{relation}
\end{equation}
The $n \times n$ determinant in (4.3) eliminates all ``diagonal"
terms with two or more indices coincident, while the statistical
factor $n!$ compensates for the overcounting of symmetrical terms.

It is next convenient to define ``generalized" overlaps 
\begin{equation}
m_\alpha^\mu(t)=
\sum_j [S_j^\mu \sigma_j(t)]^\alpha 
\end{equation}
of the current network configuration with one of the prescribed patterns,
$\alpha$ being a positive integer. 
Since $S_j^2=\sigma_j^2=1$, the quantity $m_\alpha^\mu(T)$ reduces to $K_i$ 
for $\alpha$ even and to $m^\mu(t)$ for $\alpha$ odd.  
Appealing to direct evaluation of the right-hand side of 
Eq.~(\ref{generic}) or Eq.~(\ref{relation}) for $n=1-4$,
we establish the pattern of behavior for the higher orders: 
\begin{equation}
 C_1 = \sum_{\mu=1}^p S_i^\mu [m_1^\mu] \, ,
\end{equation}
\begin{equation}
C_2 = \sum_{\mu=1}^p S_i^\mu {1\over {2!}} [(m_1^\mu)^2 - m_2^\mu] \, ,
\end{equation}
\begin{equation}
C_3 = \sum_{\mu=1}^p S_i^\mu {1\over {3!}} [(m_1^\mu)^3
- 3m_1^\mu m_2^\mu + 2m_3^\mu] \, ,
\end{equation}
and
\begin{equation}
C_4 = \sum_{\mu=1}^p S_i^\mu {1\over {4!}}
[(m_1^\mu)^4  - 6(m_1^\mu)^2 m_2^\mu
+ 8m_1^\mu m_3^\mu
   + 3(m_2^\mu)^2 - 6m_4^\mu] \, .
\end{equation}
It is seen that the generic term $C_n$ is built as a sum over
all patterns of individual terms of the form
\begin{equation}
S_i^\mu {1 \over n!} \gamma(\alpha_1,\cdots,\alpha_n)
\prod_{l=1}^n (m_l^\mu)^{\alpha_l}\, ,
\end{equation}
where $\gamma(\alpha_1,\ldots,\alpha_n)$ is a statistical weight 
factor and the generalized overlaps $m_l^\mu$ enter with positive
integral powers satisfying the partitioning condition
\begin{equation}
\sum_{l=1}^n l\alpha_l = n \, .
\label{partitioning}
\end{equation}
The statistical factor is found to obey the sum rules
\begin{equation}
\sum_{(\underline \alpha)} \gamma(\alpha_1,...,\alpha_n)   = 0 
\quad {\rm and} \quad
\sum_{(\underline \alpha)} | \gamma(\alpha_1,...,\alpha_n)|=n! \, ,
\end{equation}
and can be constructed as
\begin{equation}
\gamma(\alpha_1,...,\alpha_n) =
n!/[\prod_{l=1}^n (-1)^{\alpha_l + 1} (l^{\alpha_l})\alpha_l!] \, .
\end{equation}

Thus, for arbitrary $n$, the contribution $C_n$ can be written
explicitly as
\begin{equation}
C_n=\sum_{\mu=1}^p S_i^\mu \overline {\cal P}_n(m_1^\mu,...,m_n^\mu)
\end{equation}
where
\begin{equation}
\overline {\cal P}_n(m_1,...,m_n) = {1 \over n!}
\sum_{(\underline \alpha)}
\prod_{l=1}^n \gamma(\alpha_1,...,\alpha_n) m_l^{\alpha_l} \, .
\label{poly}
\end{equation}
The sum over ${\underline \alpha}$ in definition (\ref{poly}) extends 
only over those $n$-dimensional vectors 
$\underline \alpha = (\alpha_1,...,\alpha_n)$ 
whose components satisfy the constraint (\ref{partitioning}).
The quantity  $\overline {\cal P}_n(m_1,...,m_n)$ is identified
as a generalized Poly\`a polynomial \cite{lund} of the
symmetric group ${\cal S}_n$, with the signs $(-1)^{\alpha_l + 1}$ 
of the corresponding cyclic permutations incorporated.

For given $n$, the total number of solutions $P(n)$ of condition
(\ref{partitioning}) can be determined by induction from the
recurrence relation \cite{abramowitz}
\begin{equation}
P(n) = {1 \over n} \sum_{q=1}^n \rho (q) P(n-q) \, ,
\end{equation}
in which the divisor function $\rho(l)$ is the sum of the first powers 
of the divisors of $q$.  For large $n$, $P(n)$ behaves asymptotically 
as
\begin{equation}
P(n) = {1 \over {4n\sqrt3}} e^{ \pi \sqrt{2n \over 3} } \, .
\end{equation}

Finally, the generating function of the Poly\`a polynomials
may be employed to calculate the sum of all individual n-order 
contributions, i.e. the net internal stimulus $h_i(t)$ 
of Eq.~(\ref{netstimulus}), in limit of large connectivity
$K_i$, which is equivalent to the thermodynamic limit.  One
finds 
\begin{equation}
 \sum_{n=0}^{\infty} C_n = \sum_{\mu=1}^p S_i^\mu
{\rm exp} \left[ \sum_{l=1}^{\infty}
{(-1)^{l+1} \over l } m_l^\mu \right]  \qquad (K_i \to \infty ) \,.
\end{equation}
While this is a beautiful formal result, practical neural
network applications often work with a single fixed order
or with a few low orders adapted to the complexity of the 
problem (see, e.g. Ref.~\cite{minsky}). 

Combinatoric group-theoretical considerations reveal an interesting 
one-to-one correspondence between the $n$th-order contribution $C_n$ 
to the stimulus sum (\ref{netstimulus}) and the sum of planar 
$n$-particle cluster diagrams for noninteracting particles obeying 
Fermi statistics.  (Substitution of a permanent for the determinant 
in expression (\ref{relation}) would produce a one-to-one correspondence 
with the sum of Bose $n$-body cluster diagrams.)  Each fermion cluster
diagram is uniquely defined by an $n$-dimensional vector 
$(\alpha_1,...,\alpha_n)$ satisfying relation (\ref{partitioning}) 
and specifying a partitioning of the $n$-particle
cluster into sub-clusters correlated by exchange, namely into 
$\alpha_1$ $1$-cycles, $\alpha_2$ $2$-cycles, ... and $\alpha_n$ 
$n$-cycles.  The statistical weight factor $\gamma(\alpha_1,...,\alpha_n)$ 
is the number of ways in which $n$ particles can be assigned to
$\alpha_l$ exchange clusters of size $l$, with $l$ running from 
1 to $n$.  Figure 1 shows all possible cluster diagrams up to order 
$n=7$.  Each contribution diagram consists of $n$ filled dots and the 
associated exchange lines.  Reflecting the Fermi (or Bose) symmetry 
of the wave function, the exchange lines only occur in closed loops: the 
particles belonging to a given exchange cluster appear
as nodes in a continuous circuit of lines that represents a 
transposition or cyclic permutation.  Cluster diagrams of this 
type (though with additional lines representing dynamical 
correlations) are used in the description of non-interacting 
fermions or bosons in the correlated wave-function and
correlated density-matrix formalisms \cite{clarkprog,ristig}. 

A large number of computer experiments \cite{kuerten3} have established 
the following behavior of higher-order networks when applied to 
problems in pattern recognition.  When the patterns to be recognized 
are structured rather than random, the network dynamics usually 
converges to the pattern with closest {\it structural} similarity 
to the initial pattern, rather than to (or to a state very near) 
the pattern having largest overlap with the initial state.  This 
behavior contrasts with that of first-order networks having only 
binary synapses \cite{amit}; relative to these conventional systems, 
higher-order networks demonstrate a greatly enhanced capability 
for structural discrimination of arbitrarily complex patterns.  
Moreover, when functioning in the regime of dilute pattern storage 
(i.e., far from saturation, thus $p \sim N << N^K$, $K \geq 2$), 
the basins of attraction of the memorized patterns are dramatically 
enlarged.  Finally, it is to be emphasized that in the model 
we have considered, the combinatoric explosion of weight 
coefficients is obviated, since the network only needs to
know the overlaps of the present state with all the patterns
to be embedded.

This paper is a contribution to the ZiF Research Year on the
Sciences of Complexity:  From Mathematics to Complexity to
a Sustainable World.  The research was supported in part
by the U.S. National Science Foundation under Grant 
No.~PHY-9900713.

\newpage
\centerline{\bf Figure Caption}
\vskip .5 cm
\noindent
{\bf Fig.~1.}
All possible fermion cluster diagrams for $n=2,3,...,7$, in the
absence of dynamical correlations.

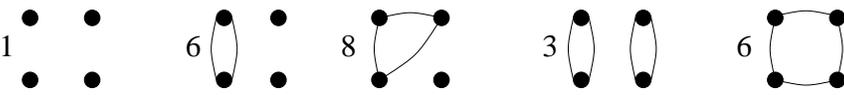
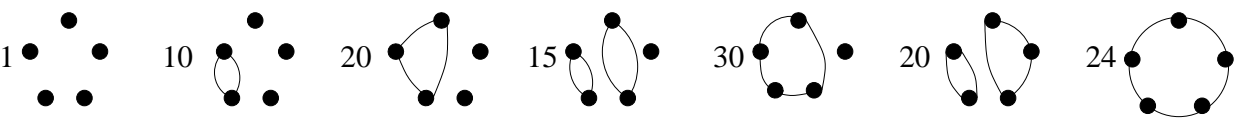
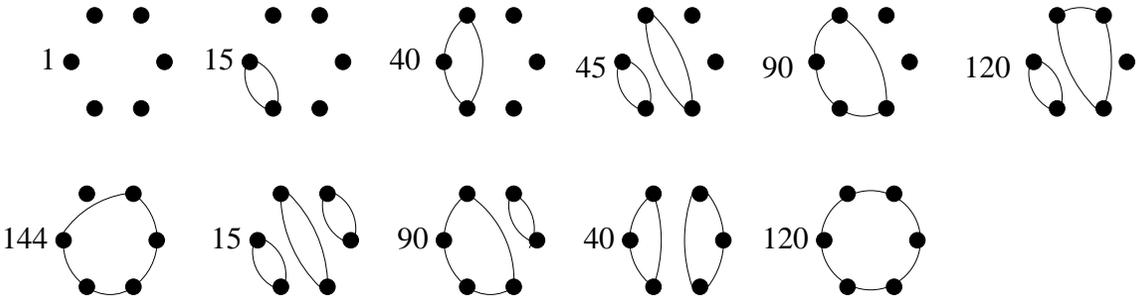
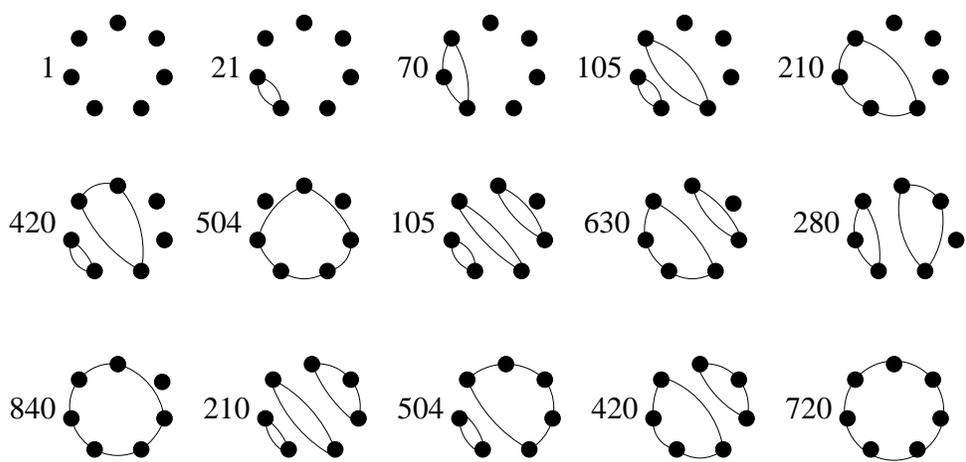
\end{document}